%% file: ms.tex
\documentclass[iop]{emulateapj}

\usepackage{xspace}
\usepackage{amssymb}

\input{localdefs}

\shorttitle{\tgcat: \evl X-ray Flares}
\shortauthors{Huenemoerder et al.}

\begin{document}

\title{X-ray Flares of \evl: Statistics, Spectra, Diagnostics}

\author{David P.\ Huenemoerder\altaffilmark{a},
Norbert S.\ Schulz\altaffilmark{a},
Paola Testa\altaffilmark{b},
Jeremy J.\ Drake\altaffilmark{b},
Rachel A.\ Osten\altaffilmark{c},
\& Fabio Reale\altaffilmark{d}
\ \\
}

\altaffiltext{a}
{Massachusetts Institute of Technology,
  Kavli Institute for Astrophysics and Space Research,
  77 Massachusetts Avenue, 
  Cambridge, MA, 02139
  (MS: NE80-6065)
}   

\altaffiltext{b}
{Harvard-Smithsonian Center for Astrophysics,
 60 Garden St., 
 Cambridge, MA, 02138}   

\altaffiltext{c}
{Space Telescope Science Institute,
  3700 San Martin Drive,
 Baltimore, MD, 21218}   

\altaffiltext{d}
{Universita di Palermo, 
  Dipartimento di Scienze Fisiche \& Astronomiche,
  Piazza del Parlamento 1,
  90134 Palermo, Italy}

%
\begin{abstract}
  We study the spectral and temporal behavior of X-ray flares from the
  active M-dwarf \evl in 200 ks of exposure with the \chan/\hetgs.  We
  derive flare parameters by fitting an empirical function which
  characterizes the amplitude, shape, and scale. The flares range from
  very short ($<1\ks$) to long ($\sim10^4\,\mathrm{s}$) duration
  events with a range of shapes and amplitudes for all durations.  We
  extract spectra for composite flares to study their mean evolution
  and to compare flares of different lengths. Evolution of spectral
  features in the density-temperature plane shows probable sustained
  heating. The short flares are significantly hotter than the longer
  flares.  We determined an upper limit to the Fe~K fluorescent flux,
  the best fit value being close to what is expected for compact
  loops.
\end{abstract}
%


\section{Introduction}\label{sec:intro}

Coronal activity is ubiquitous among late type stars of all classes:
pre-main sequence, main sequence, evolved binaries, and even some
single giants.  Activity on main sequence stars is generally
correlated with rotation rate except at the shortest periods where
saturation occurs.  Flaring is common among the coronally active stars
(as defined by high X-ray band luminosity relative to the bolometric
luminosity, $\log(L_\mathrm{x}/L_\mathrm{bol})\sim -4$ to $-3$).
Nearly every sufficiently long (many kiloseconds) X-ray observation of
coronal sources shows flares, as defined by a rapid increase in flux
accompanied by a hardening of the spectrum and subsequent decay and
softening.  Flares are the most dynamic aspects of coronal activity
and are possibly a significant source of coronal heating.  From Solar
studies, we know that flare mechanisms involve magnetic field
reconnection, particle beams, chromospheric evaporation, rapid bulk
flows, mass ejection, and heating of plasma confined in loops.

Many stellar activity studies have attempted to avoid flares in order
to determine quiescent coronal properties such as emission measure
distributions, elemental abundances, loop heights, and geometric
distribution of active regions.  Since flares are a phenomenon of
magnetic reconnection thought to occur on small spatial scales,
modeling the dynamic behavior allows us to constrain loop properties
in ways that cannot be done from analysis of quiescent coronae that
necessarily require a spatial and temporal average over a large
ensemble of coronal structures.  Use of rotational modulation of flux
and velocity has become a common and important technique (Doppler
imaging, and related methods) for mapping the stable and non-uniform
structures of stellar activity.  Flare modeling has the potential to
become as important for determining the properties of transient loop
structures and their energetics \citep[e.g.][]{Reale:2007,
  Aschwanden:Tsiklauri:2009}. 

Here we exploit the flaring behavior in \evl --- one of the brightest,
most reliable flaring sources available --- to obtain spectra and
temporal profiles in flares from \chan X-ray Observatory (\cxo) High
Resolution Transmission Grating Spectrometer (\hetgs) observations.
The $200\ks$ exposure, flare frequency, and amplitudes are sufficient
to provide spectral diagnostics in flares if we combine multiple
events to examine the mean properties of emission line and continuum
evolution. In addition, we study the distribution of flare temporal
morphologies.  

Even though \evl has a high flare rate, we are still forced to work
with some mean quantities by combining data from multiple, possibly
physically different flares.  Nevertheless, the results are
interesting and are necessary groundwork for guiding future studies.
In this paper, we characterize the flares and present some simple
diagnostics from different flare states.  We also model Fe~K
fluorescence.  Detailed hydrodynamic models will be applied in future
work.

\subsection{Characteristics of \evl}

\begin{figure*}[!htb]
  \centering\leavevmode
  {\includegraphics[width=\columnwidth]{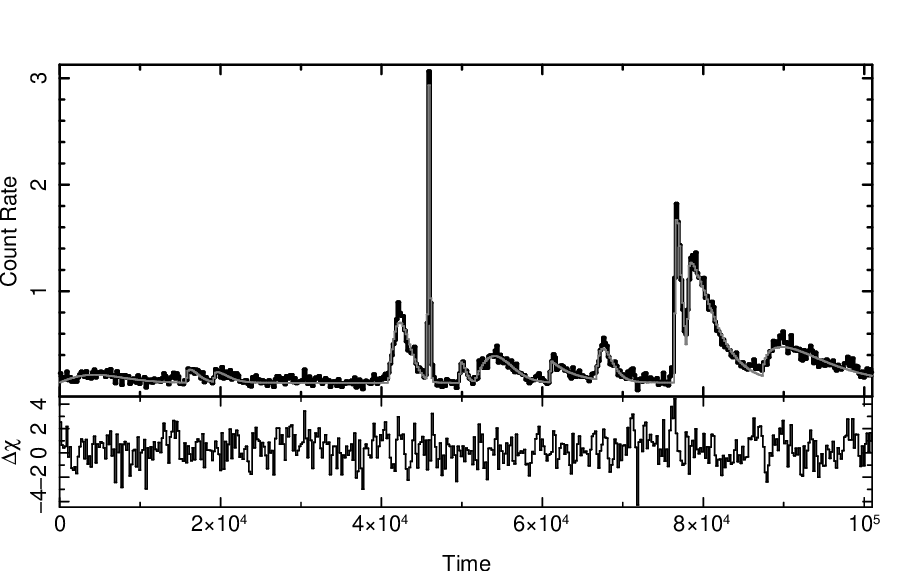}}
  {\includegraphics[width=\columnwidth]{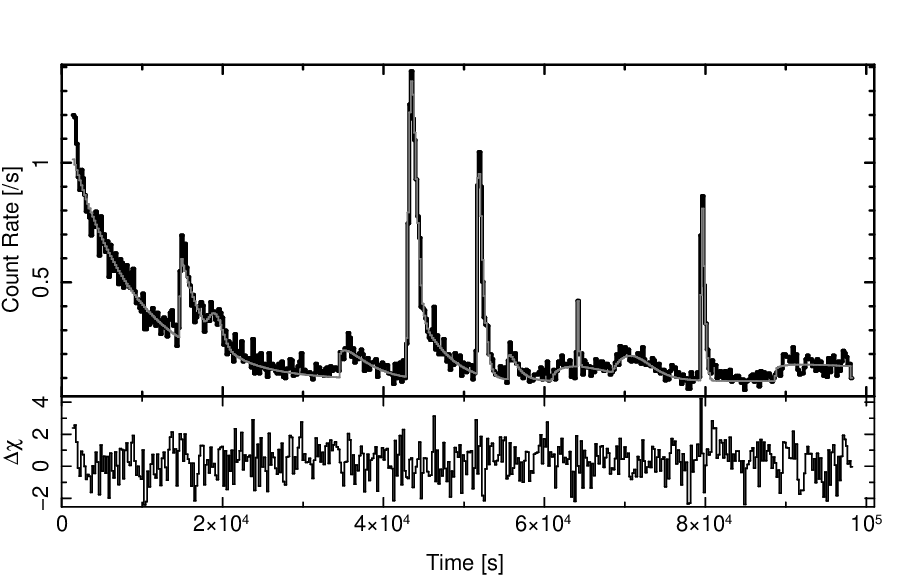}}
  \caption{These are the light curves from the two \hetgs
    observations, for \heg and \meg $1-25\mang$ events.  The top shows
    the 2001 observation (ObsID 1885), and bottom 2009 (ObsID
    10679).  It is clear that flares are frequent; 25 were identified
    and fit with an empirical function, for which we show the model
    (overlayed smooth, light-colored curve) and residuals (lower
    panels).}
  \label{fig:hetgflares}
\end{figure*}

\evl\ is a nearby (5 pc) dM3.5e ($0.35 M_\odot$, $0.36 R_\odot$) flare
star with a photometric period of 4.4 days.  It is among the X-ray
brightest of single dMe flare stars \citep{Robrade:Schmitt:2005},
having a mean $L_\mathrm{x} \sim 3-5\times10^{28}\eflux$.  It
is of particular interest for this study for these specific reasons:

{$\bullet$} \evl has consistently shown frequent and strong flares
whenever observed in X-rays, as has been demonstrated with $ROSAT$ and
$SAX$ \citep{Sciortino:Maggio:al:1999}, $ASCA$
\citep{Favata:Reale:al:2000}, $XMM/Newton$
\citep{MitraKraev:al:2005,Robrade:Schmitt:2006}, $Chandra$
\citep{Osten:Hawley:al:2005}, and {\it Suzaku}
\citep{Laming:Hwang:2009}.  Models imply that the flaring structures
are of a compact nature.  Such a compact geometry is more efficient
for producing Fe~K fluorescent emissions due to the larger solid angle
seen by the photosphere, and also because lower loop heights have a
greater the yield of Fe~K photons from the fluorescing photosphere ---
more hard X-ray photons enter at small angle and have a lower optical
depth for escape \citep{Bai:1979}. The X-ray flare frequency is about
$0.4\,\mathrm{hr^{-1}}$ (see Figure~\ref{fig:hetgflares}), similar to
previously quoted rates of $0.2-0.3\,\mathrm{hr^{-1}}$ for X-ray, UV,
and U-band flares \citep{Osten:Hawley:al:2005, MitraKraev:al:2005}.
Some of the flares are very short ($\leq1\,\mathrm{ks}$;
Figure~\ref{fig:hetgflares} and \citet{Osten:Hawley:al:2005}).  The
short duration attests to a compact size since larger volumes have
longer cooling times due to their lower density which reduces the
radiative and conductive loss rates.  With the addition of the new
\hetg observations, we have identified 25 flares in $200\ks$
(Figure~\ref{fig:hetgflares}), with a broad range in amplitude and
timescales.

{$\bullet$} \evl's \eli{O}{7} triplet ratio showed a density of about
$6\times10^{10}\,\mathrm{cm}^{-3}$
\citep{Testa:Drake:al:2004b,Ness:gudel:al:2004}.  The ability to
measure density provides strong constraints on the emitting volume.

{$\bullet$} \evl is one of 3 in a sample of 22 coronally active stars
which showed presence of opacity as determined from the ratios of
H-like Ly$\alpha$ to Ly$\beta$ resonance lines \citep{Testa:al:2007b}.
Measurement of opacity can also provide constraints on emitting
structure geometry. \citet{Testa:al:2007b} derived a compact coronal
height of about 0.06 stellar radii.

{$\bullet$} \evl has a strong and variable magnetic field ($3\kG$)
with about a 50\% surface filling factor
\citep{JohnsKrull:Valenti:1996, PhanBao:al:2006}, and it was both
poloidal and asymmetric \citep{Donati:Morin:al:2008}. Strong fields
may be necessary for forming and maintaining compact loops which give
rise to the short, energetic flares.  The star is near the expected
mass boundary for transition from an $\alpha-\omega$ dynamo to a
turbulent dynamo and so could be an important case theoretically.


\section{\chan/\hetgs Observations and Data Processing}

\chan has observed \evl twice with the \hetgs, once in 2001
(Observation Identifier 1885) for $100\ks$, and once in 2009 (ObsID
10679, as part of the \hetg Guaranteed Time Observation program) for
$97\ks$.  The spectrometer and observatory are described by
\citet{HETG:2005} and \citet{Weisskopf:02}, respectively.  The
observational data were processed with the standard software suite,
\ciao \citep{CIAO:2006} to filter, transform, bin spectra, and
construct the observation-dependent responses.  These \ciao programs
were in turn driven by the \tgcat
\citep{Mitschang:Huenemoerder:Nichols:2009pp} reprocessing
scripts\footnote{{\tt
    http://space.mit.edu/cxc/analysis/tgcat/index.html}} which
automates the \ciao processes into an end-to-end pipeline.  This was
especially useful for extraction of hundreds of spectra and responses
in time-filtered intervals (see \S\ref{sec:specex}).  Light curves
were binned from the event files using the ``ACIS Gratings Light
Curve'' package ({\tt aglc}\footnote{{\tt
    http://space.mit.edu/cxc/analysis/aglc/}}), which bins counts and
count rate light curves over multiple detector chips, grating orders,
grating types, and wavelength regions of the dispersed grating
spectrum.

\section{Flare Light Curve Fitting \& Statistics}

The Weibull distribution is a convenient function for empirically
fitting the shapes of the flares, whether impulsive or gradual.  The
normalized distribution is
\begin{eqnarray}
  f(p; a,s) =&   \left(\frac{a}{s}\right) p^{(a-1)} e^{-p^a}\label{eq:weibullone}\\
  p =& (t - t_0)/s\label{eq:weibulltwo}
\end{eqnarray}
%
in which $a$ is a shape parameter ($a>0$), the decay scale (or width
of the distribution) is specified by $s$ ($s > 0$), the offset is
given by $t_0$ ($t_0\ge0$); the independent time coordinate is $t$
($t\ge t_0$).  Shapes $a\le1$ are exponential-like --- there is no
resolved rise.  A shape $a>1$ is more Gaussian-like, becoming more
symmetric as $a$ increases.  For $a<1$, the function falls faster than
an exponential, and for $a>1$, it falls somewhat slower than an
exponential, up to a factor of about 1.5, as defined by the
$e$-folding time from the peak rate.

Light curves were binned to $100\,\mathrm{s}$ using events from
positive and negative first orders, \heg and \meg, from
$1.5$--$25\,\mang$.  These uniformly-binned light curves were fit
iteratively using a sum of Weibull distributions scaled by an
amplitude parameter, and including a constant basal count rate; in
other words, we fit $A f(p;a,s) + R_0$ in which $A$ is the flare area
in counts, $R_0$ the baseline rate in $\cts$ for the observation, and
$f$ is given by equations~\ref{eq:weibullone}-\ref{eq:weibulltwo}.  We
first defined model components for obvious flares, fitting the region
of significantly overlapping flares, and then added components as
required to flatten the residuals.  Since we physically expect
single-loop flares undergoing radiative decay to be exponential in
time, or likely prolonged by sustained heating, we constrained the
shape, $a$, to be $\ge1$ so that the decay is exponential or slower.

\input{tbl_flare_params-pp} 

Table~\ref{tbl:flareparams} gives the flare parameters.  The first
column, $n$, is an arbitrary index used as a unique identifier for
each flare (with no significance to the order).  The second column,
$area$, is the overall normalization, or amplitude, of the flare in
counts, integrated over $1$--$25\mang$.  The next three columns (3--5)
give the Weibull distrubution's parameters, $t_0$, the flare start time
(measured from the start of the observation), the shape parameter,
$a$, and the scale parameter, $s$.  The values in parentheses are the
$90\%$ confidence intervals of the parameters.

\begin{figure}[!htb]
  \centering\leavevmode
  {\includegraphics[width=0.95\columnwidth]{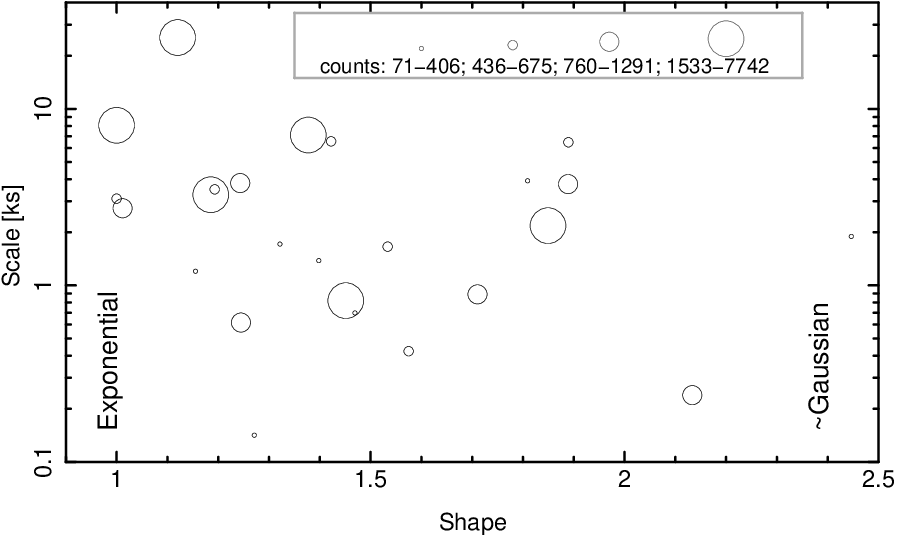}}
  \caption{Flare fit parameters. These specify the model used in
    Figure~\ref{fig:hetgflares} and listed in
    Table~\ref{tbl:flareparams}.  A $shape$ ($a$) of 1 means
    exponential decay, while larger shapes have resolved rises.  The
    $scale$ ($s$) is approximately the $e$-folding time from the peak
    rate.  Circles' radii are proportional to the $\log$ of the flare
    counts, grouped in quartiles (the inset key gives actual ranges in
    counts).  There is no obvious correlation among these parameters.}
  \label{fig:flareparams}
\end{figure}

The last three columns (6--8) are not independent parameters, but are
given as an alternate and more familiar characterization of the
flares.  They are the $e$-folding time from the peak, $\tau_1$, the
lag from the start to the peak, $\tau_2$, and the count rate at the
peak, $r$.  The horizontal line midway down the table separates the
2001 observation ($n>100$) from that of 2009.
Figure~\ref{fig:flareparams} shows the shapes, scales, and areas
graphically.  The baseline count rates, $R_0$ were slightly different
for the two observations, being $0.151\,(0.004)$ in 2001, and $0.092\,
(0.002)$ in 2009 (uncertainties are for 90\% confidence).

\begin{figure}[!htb]
  \centering\leavevmode
  {\includegraphics[width=0.95\columnwidth]{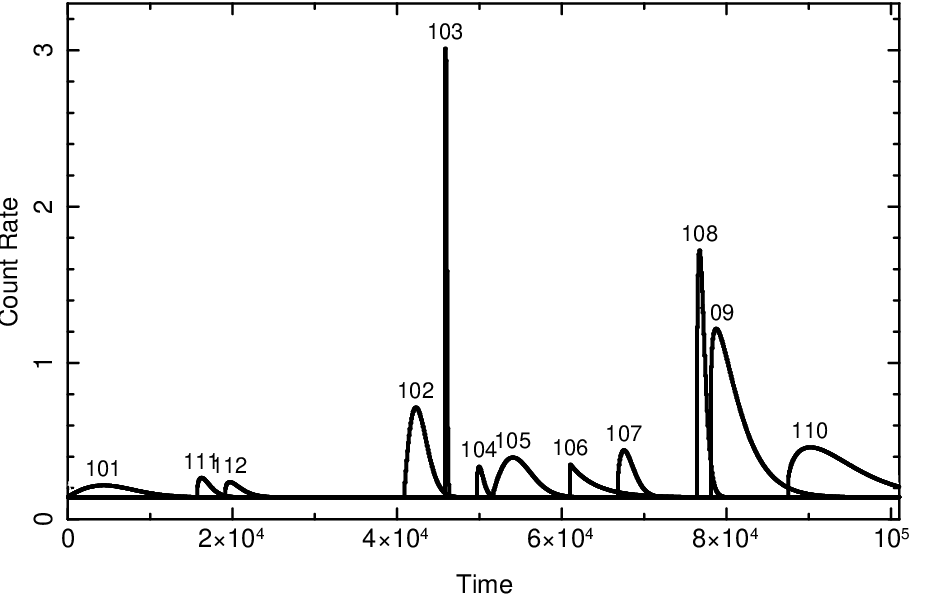}}
  {\includegraphics[width=0.95\columnwidth]{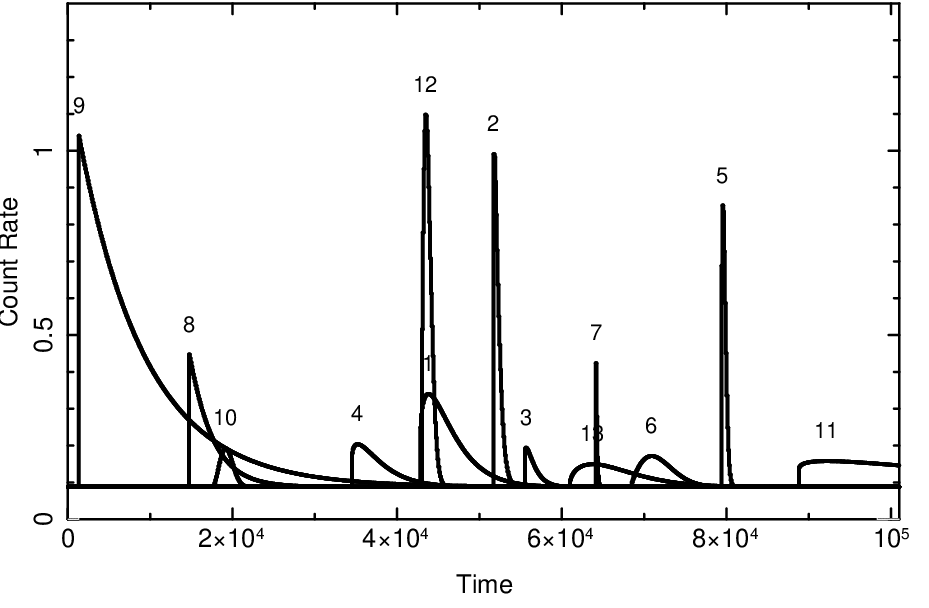}}
  \caption{These are the model components for each flare, with the
    index, $n$, as given in Table~\ref{tbl:flareparams}.}
  \label{fig:flareids}
\end{figure}

\section{Flare Spectral Profile Extraction \& Fitting}\label{sec:specex}

In order to get enough counts in different flare states, we need to
extract spectra as a function of time during flares, then form
composite spectra from states deemed similar.  Time ranges and number
of spectra to extract were identified manually from the light curves,
using higher time resolution during intervals of more rapid change.
Given a list of start times, stop times, and number of spectra,
good-time-interval (GTI) tables were constructed which were then used
to filter the events and extract spectra. We extracted 121 spectra for
the 2001 observation, and 118 for 2009.  Processing was automated
using the {\tt tgcat} scripts so that each spectrum also has the
appropriate response for its time interval.

\begin{figure}[!htb]
  \centering\leavevmode
  {\includegraphics[width=0.75\columnwidth]{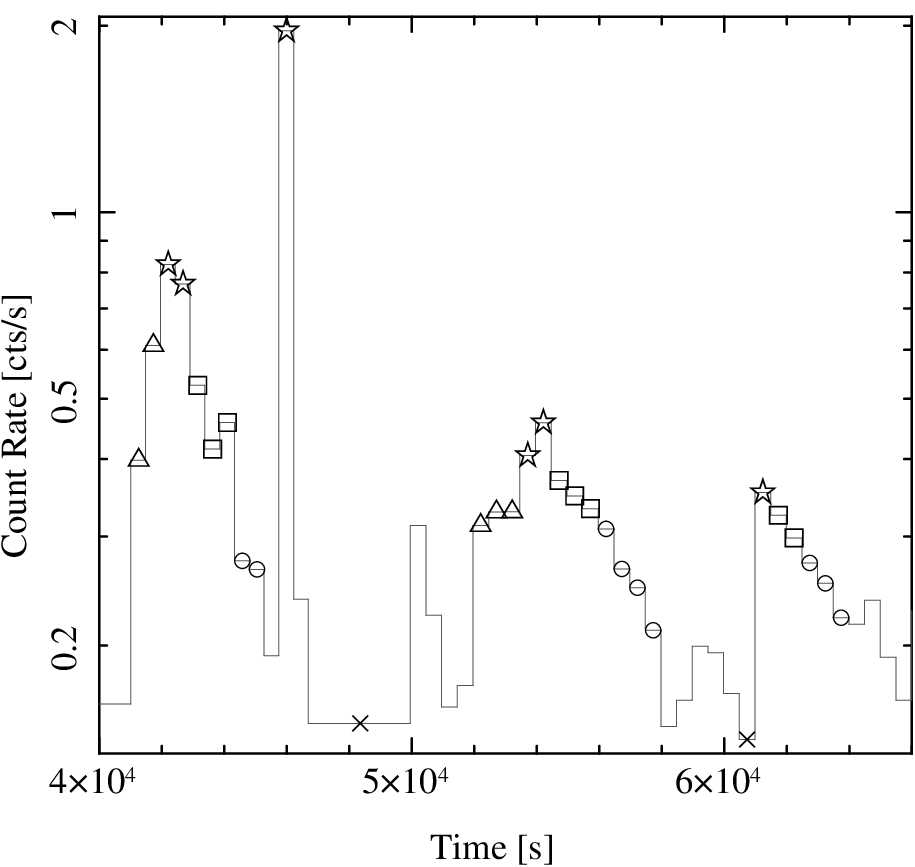}}
  \caption{Here we show detailed evolution for flares 104--106, events
    with resolved rise and decay, no resolved rise, or only the
    unresolved peak.  We define the flare ``phases'' as the rise
    (triangles), peaks (stars), decay-high (squares), decay-low
    (circles), and low (crosses).}
  \label{fig:fstates}
\end{figure}

Given the extractions, the flare phase of each time bin was then defined as
``low'', ``rise'', ``peak'', ``decay high'', ``decay low'', or
``ignore''.  Of the 249 total spectra extracted (each comprised of
\heg and \meg positive and negative first orders), 109 were assigned
to the noticed groups, the ignored remainder being of ambiguous phase
due to overlapping flares of comparable magnitude.
Table~\ref{tbl:flarestatestats} gives some summary statistics on the
selections.  Since the rises are often very sharp, we have the
shortest exposure and fewest counts in this group.  The longest time
bin, for the low phase, was $17\ks$, while during flares we used bins
as short as $300\,\mathrm{s}$.  In Figure~\ref{fig:fstates} we show
detail of some of the flare phase selections.

\begin{table}[!htb]
  \caption{Flare Phase Information}
  \begin{tabular}{lrrr}
    Phase& $N$& $t_\mathrm{exp}$& Counts\\
    &    & [ks]&\\\hline
    low&       13&  50&  7235\\
    rise&      15&   9&	 3036\\
    peak&      21&  10&	 8133\\
    decay-high&  26&  18&	10585\\
    decay-low&  34&  22&	 7369\\\hline
    \multicolumn{4}{p{2.0in}}{$N$ is the number of spectra extracted in each
      flare phase.}\\\hline
    \label{tbl:flarestatestats}
  \end{tabular}
\end{table}

The spectral analysis fitting was done by loading the individual
spectral counts histograms and their associated response files into
the \isis\citep{Houck:2002, Houck:00} analysis system where they could
then be associated and combined dynamically during fitting.  Spectral
grouping was also done dynamically as appropriate to obtain sufficient
statistics per wavelength bin.

\section{Flare Modeling}\label{sec:flaremodeling}

We study the flares by applying different model approaches to derive
diagnostics at different levels, using Solar flare models as a basis.
The numerical models have become very sophisticated; codes and methods
have been tested in detail against spatially resolved solar
observations \citep[see for examples,][]{Reale:2007,
  Klimchuk:Patsourakos:Cargill:2008, Aschwanden:Tsiklauri:2009}.  In
application of these models to \evl flares coupled with spectral
diagnostics, we can potentially constrain flare loop conditions in
this M-dwarf and relate it to the Sun and other stars.

Many solar X-ray flares are characterized by simple rise plus decay
light curves and typically involve localized loop structures, where
the plasma is confined by the coronal magnetic field. It is then
possible to describe the flaring plasma as a compressible fluid which
moves and transports energy along the magnetic field lines and use
time-dependent hydrodynamic models.  In these conditions, it has been
shown that the flare decay time scales with the length of the flaring
loop \citep{Serio:Reale:al:1991} and the presence of significant
heating in the decay can make the decay longer than expected
\citep{Jakimiec:Sylwester:al:1992,Sylwester:Sylwester:al:1993}.  Such
heating can be diagnosed from the analysis of the flare path in the
density-temperature diagram: the flare decay path becomes shallower.
From numerical modeling, scaling laws have been derived to estimate
the loop length after correcting for the effect of heating
\citep{Reale:Betta:al:1997}: $L_9 = C\, \tau_\mathrm{x} \sqrt{T_7}$,
where $L_9$ is the loop half length ($10^9\,\mathrm{cm}$),
$\tau_\mathrm{x}$ the decay time of the X-ray light curve in seconds,
$T_7$ is the maximum flare temperature ($10^7\,\mathrm{K}$), and $C$ a
proportionality factor.  Such a diagnostic tool is well established
and has been commonly applied to analyze stellar flares
\citep{Reale:Micela:1998,Favata:Schmitt:1999,MaggioPallavicini:2000,
  Stelzer:al:2002,Briggs:Pye:2003,Pillitteri:Micela:2005,Favata:Flaccomio:al:2005}.
Without heating, the scale factor $C = 1/120$
\citep{Serio:Reale:al:1991}, but the scaling law can greatly
overestimate the loop length.  More recently, new diagnostic tools
have been developed to obtain information on the flaring plasma and
structures from the flare rise phase \citep{Reale:2007}. It has been
also shown that flares with more complex light curves, e.g. with
multiple peaks, can involve more coronal loops
\citep{Reale:Gudel:al:2004}.

\begin{figure}[!htb]
  \centering\leavevmode
  {\includegraphics[width=0.95\columnwidth]{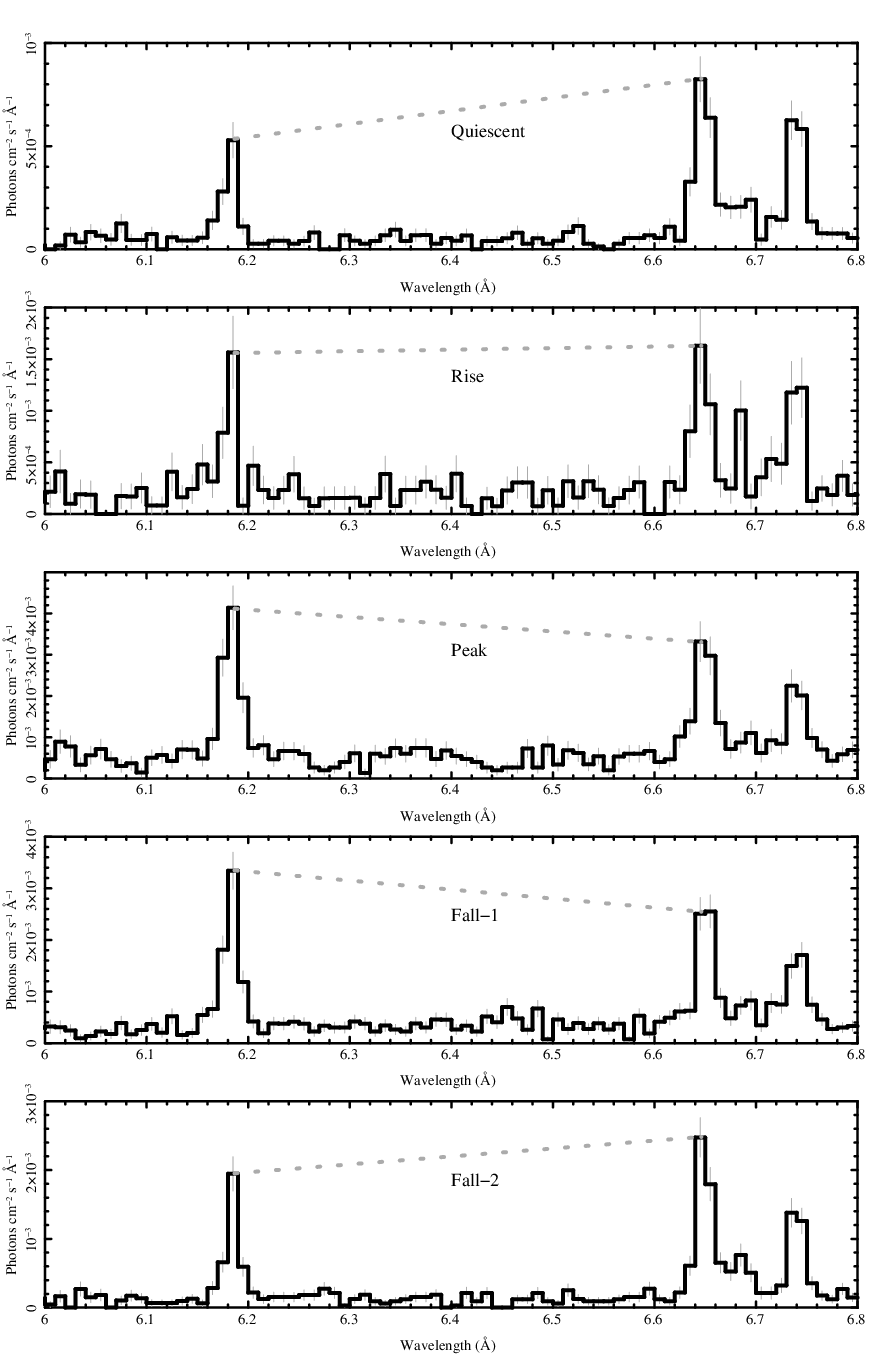}}
  \caption{This figure shows the evolution of the He-like
    \eli{Si}{14} ($6.18\mang$) to \eli{Si}{13} ($6.65\mang$) line
  fluxes through a composite flare.  Note that the $y$-axes have very
  different scales.}
  \label{fig:specsi}
\end{figure}

\begin{figure}[!h]
    \centering\leavevmode
    {\includegraphics[width=0.85\columnwidth,angle=0]{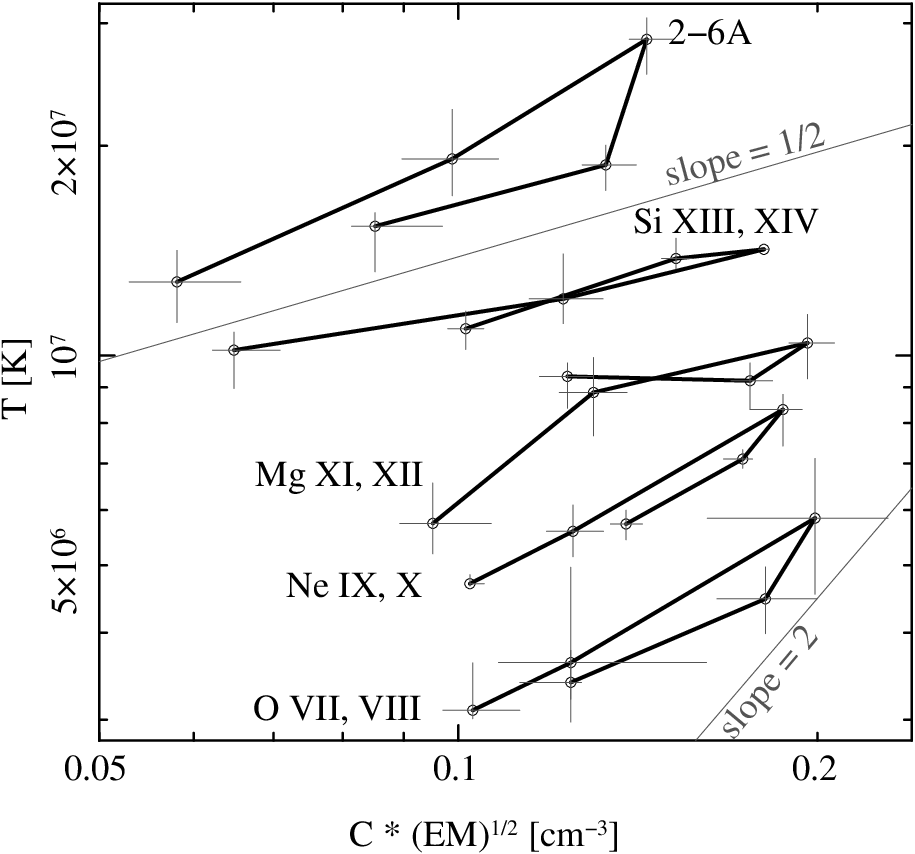}}
    \caption{Evolution of temperature and density proxy using several
      line pairs or a continuum band for the composite flare phase
      selections.  The line flux and temperature diagram traces the
      evolution of a flare, using composite spectra from different
      phases of many flares.  The 5 points on each curve correspond to
      the different flare phases, as defined in
      Table~\ref{tbl:flarestatestats}, starting from the low phase at
      the left-most point in each case and proceeding through
      ``rise'', ``peak'', ``decay-high'', and ``decay-low''.  When
      combined with time-dependent hydrodynamical flare models, the
      slopes in the different phases can be related to flare loop
      sizes and heating functions.  We show two limits as gray lines
      --- a slope of 2 is characteristic of radiative and conductive
      decay, and a slope of $1/2$ of quasi-steady-state
      decay. Errorbars are the 90\% confidence limits from the
      single-temperature APED fits to each spectral region.}
    \label{fig:temlocus}
\end{figure}

We are limited by statistics to composites of heterogeneous flare
types.  Nevertheless, the behavior agrees in general with theoretical
expectations.  Since no single flare is strong enough for this
analysis, we combined spectra from several flares and provide
diagnostics for both lines and continua.  One theoretically
interesting diagnostic is the evolution of temperature and density (or
its proxy, $\sqrt{EM}$).  The helium-like to hydrogen-like line ratios
are a strong function of temperature.  From a portion of the spectrum
which includes a line pair, we can derive a characteristic temperature
and emission measure from a relatively simple, single-temperature APED
model fit.  Figure~\ref{fig:specsi} shows the
\eli{Si}{14}--\eli{Si}{13} region.
We have derived temperatures from fits to several H-like to He-like
line pairs and a $3 - 6\mang$ continuum band using APED emissivities.
Figure~\ref{fig:temlocus} shows the results for many flares in the
existing data, grouped by flare phase as given in
Table~\ref{tbl:flarestatestats}.  These agree qualitatively with paths
expected from single-loop theory \citep{Reale:2007}.  The initial
low-phase is the left-most point for each feature, and flare phase (or
time, if it were a single flare event) proceeds (generally) clockwise
around the loop.  This is similar to results of \citet{Testa:al:2007a}
on a single, large flare in the active giant, HR~9024.
Figure~\ref{fig:temlocus} also shows two arbitrarily placed lines, one
of slope 2, and the other of slope $1/2$.  The former is
characteristic of radiative and conductive decay, and the latter of
quasi-steady-state conditions in which the heating changes slowly
enough that the temperature and density can obtain their equilibrium
values in a hydrostatic magnetic ``RTV'' \citep{RTV} loop.  (For
details on such hydrodynamic model trajectories, see
\citet{Jakimiec:Sylwester:al:1992, Reale:2007,
  Sylwester:Sylwester:al:1993}.)  The measurements shown for \evl
generally have slopes between these limits, implying that our flare
composite has some sustained heating.  The fact that some of the lines
cross (e.g., for Si and Mg) likely means that we have mixed flares of
very different physical characteristics, and that we require finer
grouping by flare types (and consequently, better statistics).

The temperatures at identical phases differ between spectral features
because we are sampling from plasma with a distribution of
temperatures, and the features have different emissivity dependence
upon temperature.  In Figure~\ref{fig:temlocus}, if we were to connect
the points at identical phases, then we would have a crude emission
measure distribution with temperature (in the square root, and
vertically oriented) for each flare phase.

The naive scaling law for the range in peak temperatures seen ($T_7
\sim 0.6 - 3.0$) and broad range in decay scales ($200 -
30000\,\mathrm{s}$) would imply a huge range in $L_9$, of two orders
of magnitude, $\sim 1$--$100$.  However, we know that we have a
heterogeneous mix and that there is sustained heating, so the upper
value is clearly a gross overestimate (for reference, the stellar
radius in these units is 25).  If we consider only the shortest
flares --- which are more likely to occur in one or few loops --- then
the range is a more plausible $L_9 \sim 0.1 - 1$.

Further progress will likely require us to obtain better statistics in
groups of similar flares, in order to constrain loop parameters with
hydrodynamic models.  For instance, due to limited time resolution, we
are not certain that the temperature and density peak at the same
flare phase, as they appear to do in Figure~\ref{fig:temlocus}.  If we
had better temporally-resolved spectra for several similar flares
(shape and duration), then we could apply detailed hydrodynamic
modeling by solving accurately the time-dependent hydrodynamic plasma
equations with the aid of the Palermo-Harvard numerical code
\citep{Peres:Serio:al:1982,Betta:Peres:al:1997}. This approach allows
one to obtain deeper insight into the flaring plasma and of the
heating details, such as the detailed thermal structure and its
evolution, and information about the loop aspect
\citep{Reale:Gudel:al:2004,Favata:Flaccomio:al:2005,Testa:al:2007a}.
Such will be attempted in future work, and hopefully with a larger
dataset which supports grouping of more like events.

\subsection{Spectral Comparison of Different States}

\begin{figure*}[!htb]
    \centering\leavevmode
    {\includegraphics[width=2.15\columnwidth,angle=-90]{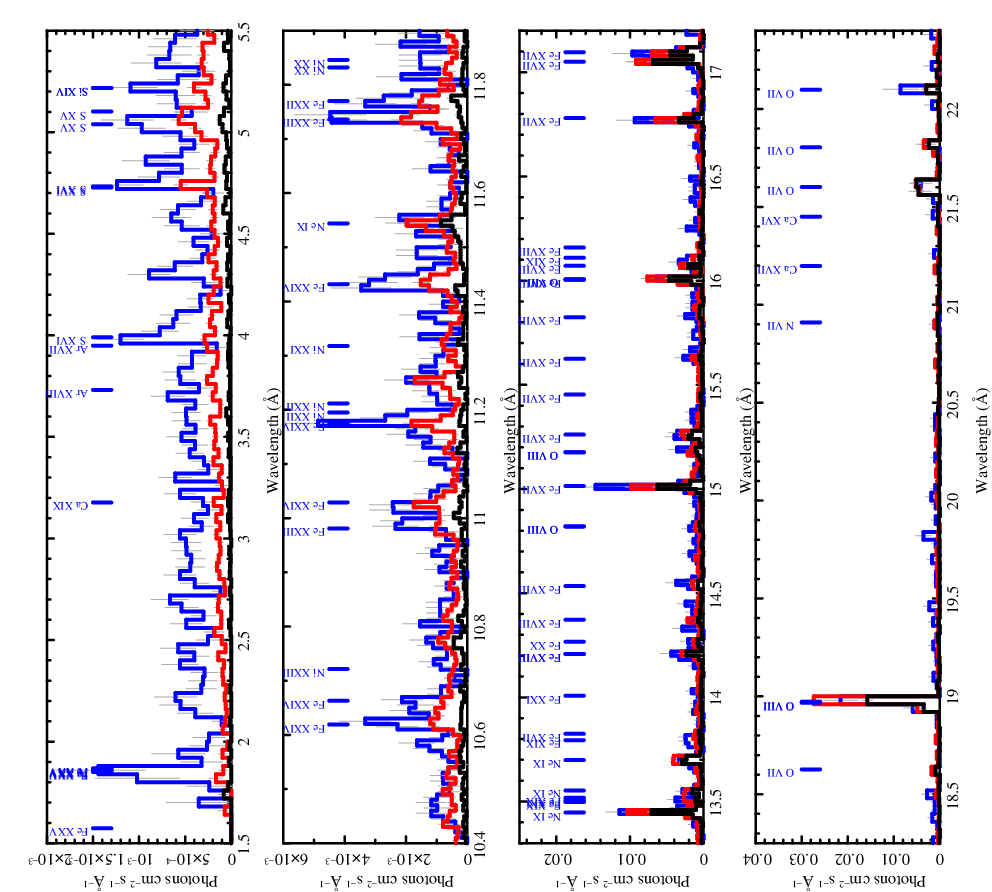}}
    \caption{Photon flux density spectra for different states in
      a few wavelength ranges: short flares (upper curves; blue), long
      flares (middle curve; red), and the low phase (lowest curve;
      black).  These are ``unfolded'' spectra, in which the effective
      area has been factored out, but the instrumental
      line-spread-function remains.  The short flare curve has more
      noise since it has the shortest exposure of $7\ks$.  The others
      have $42\ks$ (long flares) and $50\ks$ (quiescent) exposures.
      Positions of some features (not all actually present) have been
      marked.  The flare spectra have been integrated over their durations.}
    \label{fig:comparestates}
\end{figure*}

\begin{figure}[!htb]
    \centering\leavevmode
    {\includegraphics[width=0.85\columnwidth,angle=0]{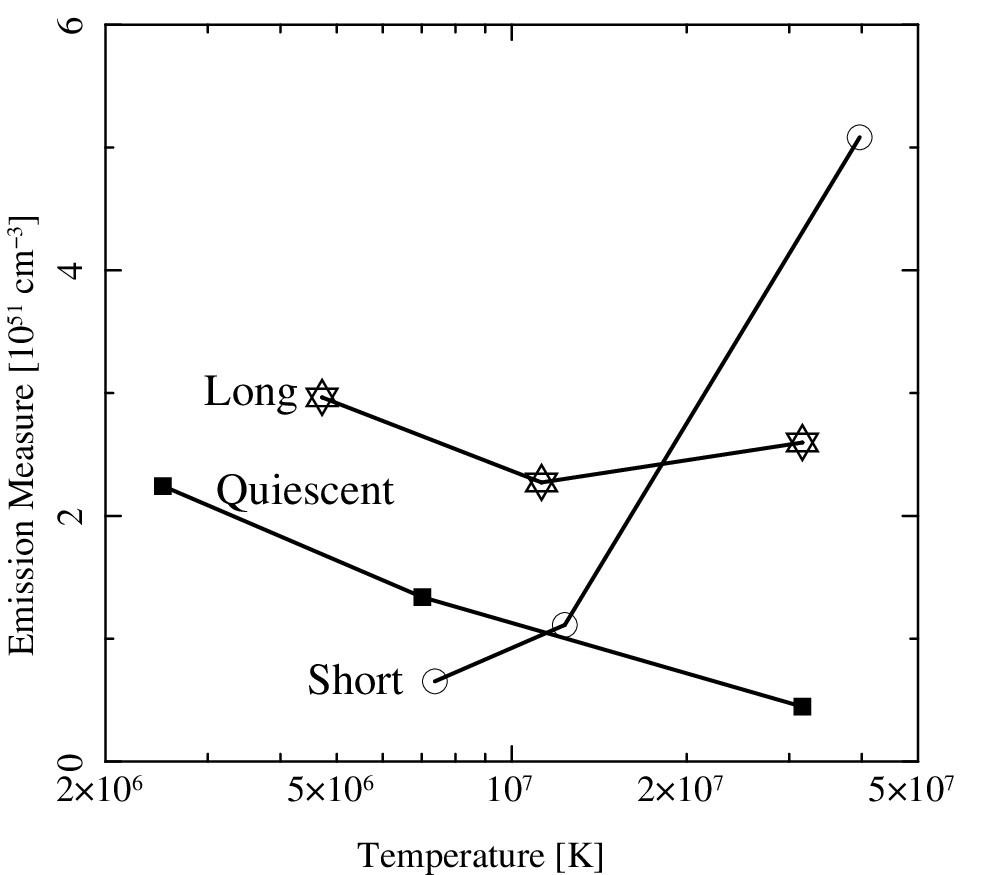}}
    \caption{The three-component APED fit parameters show the changes
      in temperature ($x$-axis) and emission measure ($y$-axis) for
      the short flares (circles), long flares (stars) and low
      phase (squares) spectra.   The short flares are clearly much
      hotter than the long.
}
    \label{fig:flarefitparams}
\end{figure}

We can empirically compare spectra for the short and long flares to
each other and to the low, or ``quiescent'' phase.  We have extracted
spectra integrated over the flare duration for short flares (flare
numbers 2, 5, 7, 12, 103, 104 and 108; see Figure~\ref{fig:flareids}),
long flares (numbers 4, 6, 8, 9, 10, 102, 105, 109, 110), and for low
phases.  In these groups, we respectively have exposures of $7\ks$,
$42\ks$, and $50\ks$.  To characterize the spectra, we have fit a
3-component APED model to each.  Such a model is not as definitive as
a line-based emission measure and abundance reconstruction. The
temperature represents some average value since the line features can
change dramatically over temperature differences of order of a factor
of 2, which a 3-temperature model cannot fairly represent over the
broad range of plasma temperatures.  Also, the abundances are largely
another line-strength parameter which can compensate for off-nominal
temperatures as well as real abundance values; we do not consider
abundance to be an interesting parameter for this purpose
(determination of meaningful abundances requires line-based emission
measure reconstruction).  Nevertheless, the model is sufficient to
show the primary differences between the phases.  The short wavelength
continuum ($<6\mang$) constrains the highest temperature component,
while the lines, primarily the H-like and He-like series, constrain
the lower temperature components.  Some lines, such as \eli{Fe}{24}
can also have a significant effect at the high temperatures of flares.

In Figure~\ref{fig:comparestates}, we show the three different spectra
for different wavelength intervals.  The upper (at the shorter
wavelengths) curve (blue) shows the short flares' peak phase.  It is
somewhat noisier than the other two curves due to its lower exposure,
but is systematically stronger and has features characteristic of high
temperature which are weaker or not present in the long flare (red
middle curve) or quiescent (black lowest curve) states.  \eli{Fe}{25}
and the \eli{S}{15}-\eli{S}{16} series are prominent during the short
flares (upper panel), as are other high ionization states of iron,
such as \eli{Fe}{23} and \eli{Fe}{24} (second panel).  At longer
wavelengths (lower two panels), where low temperature features
dominate, changes are less pronounced.

In Figure~\ref{fig:flarefitparams} we show the temperature versus
emission measure from the three-component APED fits.  It is clear that
the shortest flares have much greater weight in the hottest plasma
than the long flares.  The long flares do have some high temperature
plasma, but it does not dominate the emission measure.  Their greater
weight in cooler plasma during the decay implies that the emission
originates from a multi-loop flaring system with simultaneous cooling
and heating \citep{Reale:Bocchino:Peres:2002, Lopez-Santiago:al:2010}.
The sustained heating produces more hot plasma, which then cools and
thereby enhance the cooler region of the emission measure
distribution.  The short, hot flares probably occur in single or few
loops, and being short, do not ultimately provide as large a volume of
plasma in the cooler state.

We should note again the limitations of the three-temperature fit: the
short flare spectra do have emission from lower temperature ions such
as \eli{O}{7}, \eli{O}{8}, and \eli{Fe}{17} (see
Figure~\ref{fig:comparestates}), and so there should be some fourth
point in the Figure~\ref{fig:flarefitparams} ``Short'' curve at low
temperatures ($2$-$3\mk$) with significant emission measure.  Such
will be addressed in future work through detailed emission measure
reconstruction.

\section{Fe~K Fluorescence}

At present only the solar corona can be studied in spatial detail.
Fluorescent lines provide a powerful spectroscopic method for probing
flare geometry for the unresolved coronae of more distant stars.
X-rays of energy $\gtrsim7\,\mathrm{keV}$ emitted from a hot corona
incident on the underlying photosphere are predominantly destroyed by
photo-absorption events through inner-shell ionization of atoms or
weakly ionized species.  Observable fluorescent lines arise from
photons emitted in outward directions by the $2p-1s$ decay of the
excited atom.  In a plasma with solar-like composition, only Fe
fluorescent lines are expected to be detected
\citep{Bai:1979,Drake:Ercolano:al:2008}. The fluorescent line
equivalent width depends on the height of the X-ray source above the
photosphere and the heliocentric angle between the flare and
line-of-sight.  Line strength also depends on the {\em relative}
photospheric Fe abundance but is independent of global metallicity
except for very metal-poor stars.  For coronal scale heights
$>0.5R_\star$ the Fe~K line becomes very weak owing to geometric
dilution and the larger mean angle of incidence.  The Fe~K line is
thus potentially a very powerful diagnostic of flare geometry and
location; it can both constrain and calibrate physical models of
flares.

In the first detailed study of this kind, Fe~K fluorescence was
produced by a single large flare in the \hetgs observation of the
active giant HR~9024. \citet{Testa:al:2007a} found the origin
compatible with photospheric photospheric heights and inferred a very
compact scale height of $\leq0.1\,R_\star$.  A prominent Fe~K
fluorescent line was recently observed from a super flare on the
RS~CVn-like binary II~Peg by \citet{Osten:al:2007} who found the line
to be surprisingly strong for photospheric fluorescence.
\citet{Osten:al:2010} detected a very large flare on \evl with {\em
  Swift}, with a peak Fe~K$\alpha$ flux of
$0.027\,\mathrm{photons\,cm\mtwo s\mone}$.  Their analysis obtained
consistent results from fluorescence and hydrodynamic models, with a
compact loop of height about $0.2R_*$.

\begin{figure}[!htb]
  \centering\leavevmode
  {\includegraphics[width=0.95\columnwidth]{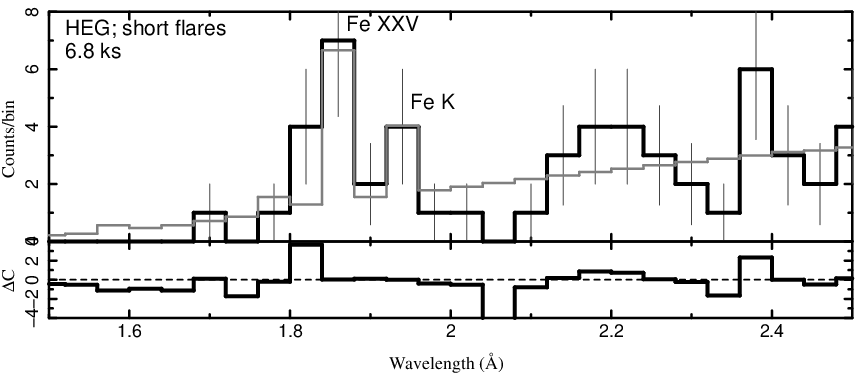}}
  \caption{Our current composite flare peak spectrum for $7\,\ks$ of
    short flares (dark histogram with light error bars) and best fit
    model (light curve).  The best fit to Fe~K gives a flux of
    $2\times10^{-5}\pflux$, and a 90\% confidence upper limit of
    $6\times10^{-5}\pflux$. }
  \label{fig:fek}
\end{figure}

Based on their rapid decay, the short flares seen from \evl are
expected to be small, single-loop events, and hence efficient
producers of Fe~K fluorescence from the underlying photosphere
\citep{Ercolano:al:2005}.  To search for this fluorescence signature,
we combined spectra from multiple short flares.  The 90\% confidence
upper limit to the Fe~K flux is about
$6\times10^{-5}\,\mathrm{phot\,cm\mtwo s\mone}$.  Given our estimate
of the $7$--$20\kev$ flux and the continuum intensity in the vicinity
of the Fe~K line, the expected equivalent width is $\sim 150$~eV for
compact flares computed by \citet{Drake:Ercolano:al:2008},
corresponding to an Fe~K flux of about $2\times
10^{-5}\,\mathrm{phot\,cm\mtwo s\mone}$.  Figure~\ref{fig:fek} shows
our fit to the flaring spectrum.  A more sensitive observation is
required in order to provide useful constraints from fluorescence on
the flare scale height.  We estimate for double the exposure in short
flares ($14\ks$), we can detect a flux of $2\times
10^{-5}\,\mathrm{phot\,cm\mtwo s\mone}$ at 90\% confidence.

\section{Conclusions}

We have shown that \evl is a reliable factory of X-ray flares, and
that they come in a large range of shapes and sizes, from time scales
of $\sim 100\,\mathrm{s}$ to over $10^4\,\mathrm{s}$, and a similar
range in integrated \hetg counts.  However, their typical brevity and
amplitude precludes study of spectral evolution for any single flare,
so we have resorted to modeling of composite flares.  The evolution of
temperature and emission measure obtained from line emission does
indicate that there is sustained heating and that simple loop model
scaling laws are probably inappropriate.  The short flares seem
qualitatively different from the long flares in their highest weight
for the hottest plasma.  If we assume that there is no sustained
heating in these flares, then the simple scaling laws would imply
longer loops due to their higher temperatures, but the short timescale
imply shorter loops.  The simple scaling law (see
\S\ref{sec:flaremodeling}) gives $L_9 \sim 3 - 21$, roughly 0.1 to 1
stellar radius, a plausible range.   Given the diversity of flare
shapes and scales, we estimate that it would take an additional
$200\ks$ observation to provide sufficient statistics in similar
moderate length flares for detailed hydrodynamic modeling of spectral
evolution, and also enough counts in short flares to provide a
positive detection of Fe~K fluorescence, and thus provide two
independent determinations of loop sizes.


\acknowledgments

\paragraph{Facilities:} \facility{\cxo (\hetgs)}

\paragraph{Acknowledgments} Support for this work was provided by the
National Aeronautics and Space Administration through the Smithsonian
Astrophysical Observatory contract SV3-73016 to MIT for Support of the
Chandra X-Ray Center and Science Instruments, which is operated by the
Smithsonian Astrophysical Observatory for and on behalf of the
National Aeronautics Space Administration under contract NAS8-03060.
JJD and PT acknowledge support from the Chandra X-ray Center NASA
contract NAS8-03060.  We thank Prof.\ Claude Canizares for granting
HETG/GTO time for this project and for comments on the manuscript.

%
\input{ms.bbl}
%

\end{document}

%% file: localdefs.tex
\usepackage{color}

\newcommand{\mtwo}  {^{-2}}
\newcommand{\mone}  {^{-1}}

\newcommand{\kev}   {\,\mathrm{keV}}

\newcommand{\ks}    {\,\mathrm{ks}}
\newcommand{\kG}    {\,\mathrm{kG}}
\newcommand{\mk}    {\,\mathrm{MK}}
\newcommand{\eflux} {\,\mathrm{ergs\,cm\mtwo\,s\mone}}
\newcommand{\pflux} {\,\mathrm{photons\,cm\mtwo\,s\mone}}

\newcommand{\mang}   {\,\mathrm{{\AA}}\xspace}
\newcommand{\evl}   {{EV~Lac}\xspace}
\newcommand{\cts}   {\,\mathrm{cts\,s\mone}}

\newcounter{ion} 
\newcommand{\eli}[2]{\setcounter{ion}{#2}#1{~\sc\roman{ion}}}

\newcommand{\chan}  {{\it Chandra}\xspace}
\newcommand{\ciao}  {{CIAO}\xspace}

\newcommand{\cxo}   {{\it CXO}\xspace}
\newcommand{\heg}   {{HEG}\xspace}
\newcommand{\hetgs} {{HETGS}\xspace}
\newcommand{\hetg}  {{HETG}\xspace}

\newcommand{\isis}  {{ISIS}\xspace}

\newcommand{\meg}   {{MEG}\xspace}

\newcommand{\tgcat} {{\em TGCat}\xspace}


%% file: tbl_flare_params-pp.tex
%
%
%
%
%
\begin{deluxetable*}{rrrrr|rrr}
  \tabletypesize{\scriptsize}
  \tablecolumns{12}
  \tablewidth{0pc}
  \tablecaption{Flare Light Curve Fit Parameters}
  \tablehead{
    \colhead{$n$}&
    \colhead{$area$}&
    \colhead{$t_0$}&
    \colhead{$a$}&
    \colhead{$s$}&
    \colhead{$\tau_1$\tablenotemark{a}}&
    \colhead{$\tau_2$\tablenotemark{a}}&
    \colhead{$r$\tablenotemark{a}}\\
    %
    \colhead{}&
    \colhead{[cts]}&
    \colhead{[s]}&
    \colhead{-}&
    \colhead{[s]}&
    \colhead{[s]}&
    \colhead{[s]}&
    \colhead{[cts/s]}\\
%
    \colhead{(1)}&
    \colhead{(2)}&
    \colhead{(3)}&
    \colhead{(4)}&
    \colhead{(5)}&
    \colhead{(6)}&
    \colhead{(7)}&
    \colhead{(8)}
  }
  \startdata
%
101&   595   (494,   713)&       0       (0,    524)&  1.89  (1.46, 2.47)&    6474    (5691,   7857)&      9818&  4346&    0.08\\
111&   234   (183,   292)&   15685   (15685,  15785)&  1.40  (1.04, 1.77)&    1383    (1047,   1833)&      2116&   563&    0.12\\
112&   226   (170,   281)&   19101   (19101,  19202)&  1.32  (1.00, 1.75)&    1715    (1304,   2144)&      2586&   588&    0.10\\
102&  1533  (1452,  1614)&   40861   (40747,  40904)&  1.85  (1.72, 2.02)&    2181    (2068,   2316)&      3320&  1432&    0.57\\
103&   854   (802,   905)&   45726   (45721,  45726)&  2.13  (2.00, 2.30)&     239     (230,    250)&       354&   177&    3.18\\
104&   184   (138,   232)&   49645   (49486,  49745)&  1.47  (1.02, 2.81)&     698     (508,    866)&      1076&   321&    0.20\\
105&  1154  (1073,  1237)&   51548   (51548,  51926)&  1.89  (1.55, 2.05)&    3752    (3358,   3954)&      5689&  2517&    0.25\\
106&   675   (554,   802)&   60934   (60898,  60998)&  1.00  (1.00, 1.13)&    3100    (2322,   4115)&      3104&     0&    0.22\\
107&   666   (586,   749)&   66725   (66642,  66825)&  1.53  (1.33, 1.72)&    1659    (1462,   1840)&      2563&   834&    0.30\\
108&  1753  (1669,  1840)&   76370   (76364,  76370)&  1.45  (1.38, 1.52)&     819     (777,    868)&      1260&   366&    1.58\\
109&  4639  (4492,  4786)&   78078   (78065,  78078)&  1.19  (1.15, 1.22)&    3264    (3140,   3397)&      4627&   682&    1.08\\
110&  3089  (2928,  3260)&   87418   (87255,  87422)&  1.38  (1.31, 1.47)&    7105    (6681,   7611)&     10839&  2776&    0.32\\
\hline%
  9&  7742  (7436,  9336)&    1305       (0,   1371)&  1.00  (1.00, 1.01)&    8071    (7652,   8529)&      8084&     0&    0.96\\
  8&  1052   (725,  1292)&   14639   (14538,  14639)&  1.01  (1.00, 1.36)&    2738    (1505,   3429)&      2912&    33&    0.36\\
 10&   200   (123,   450)&   17654   (17300,  18358)&  2.45  (1.00, 4.81)&    1895    (1082,   2855)&      2724&  1528&    0.10\\
  4&   537   (454,   645)&   34448   (34261,  34540)&  1.19  (1.01, 1.44)&    3501    (2853,   4718)&      4988&   761&    0.12\\
  1&  1291  (1140,  1416)&   42782   (42664,  42783)&  1.24  (1.15, 1.44)&    3804    (3366,   4469)&      5575&  1024&    0.25\\
 12&  1154  (1035,  1296)&   42969   (42891,  42984)&  1.71  (1.56, 1.99)&     890     (827,    994)&      1369&   533&    1.02\\
  2&   760   (702,   818)&   51628   (51601,  51628)&  1.24  (1.16, 1.33)&     616     (563,    674)&       904&   167&    0.92\\
  3&   166   (127,   232)&   55447   (55347,  55548)&  1.16  (1.00, 1.51)&    1202     (898,   2009)&      1664&   212&    0.11\\
 13&   547   (428,   691)&   60975   (60900,  61478)&  1.42  (1.09, 1.71)&    6547    (5303,   8475)&     10046&  2787&    0.06\\
  7&    71    (50,   181)&   64091   (63982,  64091)&  1.27  (1.00, 3.33)&     142     (100,    225)&       210&    42&    0.37\\
  6&   406   (304,   511)&   68413   (67939,  68814)&  1.81  (1.46, 2.47)&    3918    (3294,   4760)&      5981&  2512&    0.08\\
  5&   436   (397,   475)&   79369   (79354,  79369)&  1.57  (1.45, 1.72)&     424     (390,    460)&       655&   224&    0.78\\
 11&  2246  (1012, 63190)&   88716   (88315,  88918)&  1.12  (1.00, 1.28)&   25389   (10008, 278646)&     33882&  3459&    0.07\\
  \enddata
  \label{tbl:flareparams}
  \tablenotetext{a}{$\tau_1$ (the $e$-folding time), $\tau_2$ (time
    from $t_0$ to the peak rate), $r$ (the peak rate), are not unique
    parameters, but are derived from the preceding Weibull
    distribution's parameters.}
  \tablecomments{The central horizontal line separates the 2001 and
    2009 observations.  Parameters $t_0$, $a$, and $s$ are as defined
    by Equations~\ref{eq:weibullone}-\ref{eq:weibulltwo}, and the
    $area$ is the flare amplitude.  The index $n$ is an arbitrary
    identifier (with no significance to the order); it is used to mark
    the flares displayed in Figure~\ref{fig:flareids}. The vertical
    line separates the fitted parameters from some derived parameters
    which may be more intuitive.  Values in parentheses are $90\%$
    confidence limits.  }
%
\end{deluxetable*}